\documentclass[%
 reprint,
nofootinbib,
 amsmath,amssymb,
 aps,
]{revtex4-1}

\usepackage[export]{adjustbox}
\usepackage{color}
\usepackage{graphicx}% Include figure files
\usepackage{dcolumn}% Align table columns on decimal point
\usepackage{bm}% bold math
\usepackage{hyperref}% add hypertext capabilities
\usepackage{natbib}
\usepackage{float}
\usepackage[caption=false]{subfig}
\usepackage{hyperref}
\hypersetup{
    colorlinks=true, 
    linktoc=page,    %set to all if you want both sections and subsections linked 
    linkcolor=blue,  
    urlcolor=magenta
}

\newcommand{\be}{\nopagebreak[3]\begin{equation}}
\newcommand{\ee}{\end{equation}}

\newcommand{\bea}{\nopagebreak[3]\begin{eqnarray}}
\newcommand{\eea}{\end{eqnarray}}

\newcommand{\la}{\label}
\newcommand{\n}{\nonumber}

\newcommand{\SU}{\mathrm{SU}}

\begin{document}

\preprint{APS/123-QED}

\title{Quantum gravity predictions for black hole interior geometry}
%Effective Geometrodynamics of Schwarzschild Interior
%Quantum gravity predictions for black hole interior geometry}% Force line breaks with \\

\author{Emanuele Alesci$^1$}
\email{eza69@psu.edu}
\author{Sina Bahrami$^1$}
\email{ sina.bahrami@psu.edu}
\author{Daniele Pranzetti$^2$}
\email{dpranzetti@perimeterinstitute.ca}
\affiliation{$^1$Institute for Gravitation and the Cosmos, Pennsylvania State University, University Park, Pennsylvania 16802, USA}
\affiliation{$^2$Perimeter Institute for Theoretical Physics,  Waterloo, Ontario N2L 2Y5, Canada}

\date{\today}% It is always \today, today,

\begin{abstract}
In a previous work we derived an effective Hamiltonian constraint for the Schwarzschild geometry starting from the full loop quantum gravity Hamiltonian constraint and computing its expectation value on coherent states sharply peaked around a spherically symmetric geometry. We now use this effective Hamiltonian to study the interior region of a Schwarzschild black hole, where a homogeneous foliation is available.
%, and study its dynamical flow on the classical phase space. 
Descending from the full theory, our effective Hamiltonian, though still bearing the well known ambiguities of the quantum Hamiltonian operator, preserves all relevant information about the fundamental discreteness of quantum space. This allows us to have a uniform treatment for all quantum gravity holonomy corrections to spatially homogeneous geometries, unlike the minisuperspace loop quantization models in which the effective Hamiltonian is postulated.
%Meanwhile, it considerably reduces the ambiguities in the choice of quantum parameters entering the  effective dynamics. 
We show how, for several geometrically and physically well motivated choices of coherent states,  the classical black hole singularity is replaced by a homogeneous expanding  Universe. The resultant geometries have no significant deviations from the classical Schwarzschild geometry in the pre-bounce sub-Planckian curvature regime, evidencing the fact that large quantum effects are avoided in these models. In all cases, we find no evidence of a white hole horizon formation.
% in this effective theory. 
However, various aspects of the post-bounce  effective geometry depend on the choice of quantum states.
\end{abstract}

\maketitle

%\tableofcontents

A primary aim for the quantization 
program of  gravitational field is to 
determine the fate of classical spacetime 
singularities as predicted by general relativity. 
It has been speculated that quantum gravitational
effects smooth out spacetime singularities, analogously to
how an ultraviolet completion of quantum 
fields tames ultraviolet divergences. Fortunately, the canonical 
approach to loop quantum gravity (LQG) is now sufficiently developed
to systematically resolve this key issue from the full theory 
perspective. 

To provide some historical context, we point out that the first generation of quantum gravity predictions were made in cosmology and within the minisuperspace quantization scheme. 
 There the loop quantization program was  implemented by applying LQG inspired techniques to a (model dependent) symmetry reduced phase space of general relativity parametrized by the Ashtekar connection and densitized triad \cite{Ashtekar:2011ni}. 
% on the homogeneous isotropic sector of general relativity 
%and was subsequently generalized to the homogeneous sector.   
The resultant models, which later became known as  
loop quantum cosmology (LQC), unanimously predict that the classical
spacetime singularity is replaced by a quantum bounce \cite{Ashtekar:2006rx, Ashtekar:2007em}. They offered 
the first glimpse of how quantum gravity could resolve spacetime 
singularities.

However, attempts to generalize the loop quantization program 
to a black hole  geometry were not equally successful at  producing generally accepted predictions \cite{Modesto:2005zm, Ashtekar:2005qt, Bohmer:2007wi, Campiglia:2007pb, Brannlund:2008iw, Corichi:2015xia, Olmedo:2017lvt, Bojowald:2018xxu, BenAchour:2018khr, Ashtekar:2018lag, Ashtekar:2018cay}. The starting point of all these previous attempts was the observation that the Schwarzschild black hole interior can be described in terms of a contracting anisotropic Kantowski--Sachs model, allowing one to apply LQC techniques to this highly symmetric geometry. This approach presented two major drawbacks. 

Firstly, we were still confronted with the fundamental issue of whether the microscopic degrees of freedom removed by the symmetry reduction at the classical level could have been relevant, affecting the final physical predictions of the theory. In the case of LQC techniques applied to the Schwarzschild interior, this was clearly manifest in the use of point holonomies for the construction of the reduced kinematical Hilbert space associated with the 2-spheres that foliate the spacelike surfaces, which completely eliminated all information about the 3-D structure of the graph. 
%As we will show below, inclusion of these degrees of freedom modifies the effective dynamics in a crucial way. 
Secondly, the reintroduction, by hand, of some `elementary plaquette' at the end of the quantization procedure, in order to fix the quantum parameters entering the construction of the states, was an ambiguous prescription, with different choices yielding different physical scenarios, as well as undesirable features.
% \footnote{In this regard, the most recent proposal \cite{Ashtekar:2018lag} seems to cure most of these shortcomings; however, as noted in \cite{Bouhmadi-Lopez:2019hpp}, this choice of quantum parameters yields a solution to the effective exterior dynamics which  is not asymptotically flat.}.  

In order to construct a more robust and reliable model of quantum black hole geometry that addresses the aforementioned issues,
a new framework, the so-called `quantum reduced loop gravity' approach \cite{Alesci:2013xd, Alesci:2014rra, Alesci:2014uha,  Alesci:2016xqa, Alesci:2018ewg, Alesci:2018loi}, has recently been developed in which one starts from the full LQG theory and then performs the symmetry reduction at the quantum level by means of coherent states encoding the information about a semi-classical spherically symmetric black hole geometry. More precisely, computing the action of the quantum Hamiltonian operator on a partially gauge fixed kinematical Hilbert space
%, defined on a conveniently partially gauged fixed kinematical Hilbert space, on these states 
results in a semi-classical, effective Hamiltonian constraint $H_{\rm eff}$
that now replaces the Hamiltonian constraint of general relativity. Unlike its predecessors, the newly obtained  effective Hamiltonian constraint for the Schwarzschild geometry \cite{Alesci:2018loi}  descends from the LQG quantum
Hamiltonian constraint. It
retains all information about the fundamental degrees of freedom associated with the quantum space that is inherited from the full theory.

The effective Hamiltonian in \cite{Alesci:2018loi} was derived for a general spacetime foliation, well defined for both the exterior and the interior Schwarzschild regions. Solving for the effective geometry in both regions cannot be done within the minisuperspace quantization program as 
a homogeneous slicing that covers the entire spacetime is not available. One is forced to use an inhomogeneous slicing and subsequently work with an infinite dimensional phase space. %inhomogeneity of the foliation, which makes the standard phase space infinite dimensional, is reflected in the very complicated form of the final expression for  $H_{\rm eff}$, which turns out to be a {\it non-local} ODE. 
In particular, the effective scalar constraint equation 
$H_{\rm eff} = 0$ is now a highly non-linear, {\it non-local} ODE.
While work is in progress to develop numerical techniques to solve this equation, a simpler task is to restrict  attention to the interior region where the standard  Schwarzschild foliation is now homogeneous and the coherent states are peaked around this homogenous geometry. In this way, the effective Hamiltonian derived in  \cite{Alesci:2018loi}, as well as the Hamilton's evolution equations that it generates, assume a more tractable form which can be solved through a combination of analytical and numerical tools. In this letter we report on the results of this analysis.% and show how the classical black hole singularity is replaced by an expanding Bianchi type I Universe.

Let us emphasize that our restricting to the interior minisuperspace for the sake of solving the interior effective dynamics %originally derived in the full LQG framework 
is not equivalent to the minisuperspace quantization models of the previous literature. In fact, as it will be elucidated  in detail, our $H_{\rm eff}$ provides a uniform treatment of the
quantum gravity SU(2) holonomy corrections to spatially homogeneous geometries.
%(in our analysis, we are not including inverse volume corrections as these are sub-leading \cite{Alesci:2015nja}). 
It should be noted that a number of ambiguities present in the quantization procedure 
%affecting the construction 
for the Hamiltonian constraint  in LQG  \cite{Thiemann:1996aw} inevitably percolate in our construction as well, implying non-uniqueness for the form of the quantum corrections that we derive. However, this level of ambiguity is a direct reflection of important open issues in the full theory, unlike the quantization ambiguity in the polymer approach that is due to a choice to work with point holonomies. Therefore, our derived $H_{\rm eff}$ represents a far better motivated alternative to the previously used Hamiltonian constraint of the minisuperspace loop quantization procedure, which so far has been postulated based on the example of homogeneous and isotropic cosmology \cite{Taveras:2008ke}.
%, and it modifies the effective dynamics in a crucial way. 
At the same time, having the full theory (graph) structure to begin with, our analysis provides a consistent and intuitive geometrical framework in which we identify the quantum parameters entering the effective dynamics. This allows us to investigate, in a controlled way, how various choices of quantum coherent states result in different physical predictions of the theory.
%We will introduce it later in the letter.

%This was mainly due to the fact that it is structurally difficult to define a quantized notion of the spatial diffeomorphism constraint in loop quantum gravity, in lieu of which
%one fails to have a consistent constraint algebra and covariant dynamics.
%Nevertheless, significant progress has recently been made in the full theory loop quantization of the Hamiltonian constraint. \textcolor{red}{add a sentence or two about this; Thiemann Hamiltonian vs. what we do}. 
%In conjunction with this development, a class of coherent states were  constructed that comport to a notion of spatial homogeneity \textcolor{red}{add a sentence or two about how the states are constructed}.  

%Computing the action of the quantum Hamiltonian operator on these states results in a semi-classical, effective Hamiltonian constraint $H_{\rm eff}$
%that now replaces the Hamiltonian constraint of general relativity in its homogeneous sector. 
\vspace{0.3cm}

{\it Phase space and effective dynamics}. 
%As the interior of the event horizon for the Schwarzschild black hole
%is spatially homogeneous, we can take advantage of this simplification
%have the necessary tools 
%to investigate the fate of its singularity by solving for quantum-corrected metrics in this region of  spacetime \footnote{Quantum gravity may prevent the existence of translation Killing fields that become timelike near $\mathcal{I}$. In that case, spacetime covariance and consistency of the Hamiltonian framework requires a properly quantized spatial diffeomorphism constraint.}. 
%We restrict our search for quantum geometries to metrics in the minisuperspace, despite performing a full theory quantization of 
%the Hamiltonian constraint. Practically, this amounts to considering
%quantum gravitational effects for which the black hole interior geometry 
%continues to be spatially homogeneous. 
We are interested in an effective description for the interior geometry of a spherically symmetric black hole.
Practically, this amounts to considering
quantum gravitational effects for which the metric
continues to be spatially homogeneous. Therefore,
let us specialize to a coordinate system in which the metric is   
\be \la{4m}
 ds^2 = - N(\tau) ^2 d \tau^2+ \Lambda(\tau)^2 dx^2+ R(\tau)^2 d \Omega^2.
\ee
Here $\Lambda(\tau)$ and $R(\tau)$ are the 
two dynamical metric functions and $d \Omega^2$ 
is the unit round 2-sphere line element. Due to
spatial homogeneity, the only gauge freedom is to
rescale proper time by choosing a lapse function $N$.
As commonly done, in order to avoid having a divergent symplectic structure, we require $x\in[0,L_0]$ where $L_0$ is some infrared cut-off in the $x$ direction.
The covariant phase space of solutions consists of the metric functions $\Lambda$ and $R$ and their conjugate momenta $P_\Lambda, P_R$.
%\bea 
%&& P_R := - \frac{1}{G N} \big[\Lambda \dot{R} + \dot{\Lambda} R\big], \ \ \  P_{\Lambda} : = - \frac{R \dot{R}}{G N},
%\eea
%where dot denotes differentiation with respect to $\tau$. 
%Note that the functional dependence of $P_R$ and $P_\Lambda$ on $R$ and $\Lambda$ coincides with their classical expressions. 
The symplectic 2-form for this phase space is $
\omega=L_0(dR \wedge dP_R +  d \Lambda \wedge dP_\Lambda)$,
whence the Poisson brackets assume the familiar form $\{R, P_R\}=\{\Lambda, P_\Lambda\}=1/L_0$. To avoid confusion with the previous LQG literature, we emphasize that 
%we are not going to reabsorb 
the $L_0$ factor appearing in the symplectic structure is not  absorbed in the phase space variables. Hence, in all subsequent equations, the phase space variables must be regarded as invariants under a rescaling by this cut-off. In particular, as it will become clear below, the effective dynamics is $L_0$-independent and the undesirable features related to this scale dependence that appear in some of the previous minisuperspace models are absent in our treatment.

In the case of a Schwarzschild spacetime with a gravitational mass $m$, the above phase space variables assume the following trajectories (in the rest of the letter we denote classical solutions with a subscript $c$, and we work in $c = \hbar = 1$ units unless otherwise stated)
%\bea \label{solc}
%&& R_c(\tau) = 2 G m \ e^{\tau/2 G m}, P_{R_c}(\tau) = \frac{1}{2 G} \Big[2- e^{- \frac{\tau}{2 G m}}\Big], \\
%&&\Lambda_c(\tau) = \sqrt{e^{- \frac{\tau}{2 G m}} -1},  P_{\Lambda_c}(\tau) = -2 m \ e^{\frac{\tau}{4 G m}} \sqrt{1- e^{\frac{\tau}{2 G m}}},\n
%\eea
\bea \label{solc}
&& R_c(\tau) = 2 G m \ e^{\tau/2 G m}, \nonumber\\
&&\Lambda_c(\tau) = \sqrt{e^{- \tau/2 G m} -1}, \nonumber\\
&&P_{R_c}(\tau) = \frac{1}{2 G} \Big[2- e^{- \tau/2 G m}\Big], \nonumber\\
&& P_{\Lambda_c}(\tau) = -2 m \ e^{\tau/ 4 G m} \sqrt{1- e^{\tau/2 G m}},
\eea
for  %$N_c = - {R^2/(2 G^2 m P_\Lambda)}$
\be \label{Nc}
N_c = - \frac{R^2}{2 G^2 m P_\Lambda}
\ee
as the choice for the lapse function. Here the range $-\infty <\tau <0$ covers the entire interior region of the Schwarzschild black hole, with $\tau=0$ corresponding to the horizon and $\tau = - \infty$ to the classical singularity. 
%As we will see shortly, a convenient choice for the lapse function that we utilize in the effective theory reduces to \eqref{Nc} in the limit $\hbar \rightarrow 0$. 
%Our approach to finding the quantum-corrected 
%interior geometry is to integrate the corresponding Hamiltonian
%system. Here 
In the effective theory, the phase space variables are evolved in time $\tau$ via the  following Poisson brackets:
% $ \dot{R}=\partial H_{\rm eff}[N]/L_0 \partial P_R,  \dot{P}_R =-\partial H_{\rm eff}[N]/L_0\partial R$ and similarly for the $\Lambda$ sector.
\bea \la{peq}
&& \dot{R} = \{R, H_{\rm eff} [N]\} = \frac{1}{L_0}\frac{\partial H_{\rm eff}[N]}{\partial P_R},\n \\
&&\dot{P}_R = \{P_R, H_{\rm eff}[N]\} = -  \frac{1}{L_0}\frac{\partial H_{\rm eff}[N]}{\partial R}, \n \\
&& \dot{\Lambda} = \{\Lambda, H_{\rm eff}[N]\} =  \frac{1}{L_0}\frac{\partial H_{\rm eff}[N]}{\partial P_\Lambda},\n \\
&&\dot{P}_\Lambda = \{P_\Lambda, H_{\rm eff}[N]\} = -  \frac{1}{L_0}\frac{\partial H_{\rm eff}[N]}{\partial \Lambda},
\eea
where dot denotes differentiation with respect to $\tau$ and $H_{\rm eff}[N] $ is the smearing of the effective Hamiltonian constraint by the lapse function. As gravity is described by a constrained Hamiltonian system, we also have
$ 
H_{\rm eff} = 0
$
at all times. This constraint equation is implied by the evolution equations, provided that it is satisfied at some initial time.

In order to find the solutions to the evolution equations, let us introduce 
 $H_{\rm eff}$ and discuss in some detail the graph structure that 
we have adapted to constant $\tau$ slices. By specializing the general effective Hamiltonian derived in \cite{Alesci:2018loi} to the spatially homogeneous metric \eqref{4m} and integrating over the angular coordinates (amounting to averaging out  quantum fluctuations around spherical symmetry),
%\footnote{Integrating over the angular coordinates guarantees that the spatial isometries are preserved by the effective metric. It amounts to averaging all quantum fluctuations around spherical symmetry.}
we obtain
\bea \label{heff}
&&H_{\rm eff} = - \frac{L_0}{4 \gamma^2 G \epsilon_x \epsilon^2} \bigg[ \epsilon R \sin{\Big(\frac{\gamma G \epsilon_x [P_R R - P_\Lambda \Lambda]}{R^2}\Big)} \\
&& \ \times \bigg\{ 2 \sin{\Big(\frac{\gamma G \epsilon P_\Lambda}{R}\Big)} + \pi H_0 \Big(\frac{\gamma G \epsilon P_\Lambda}{R}\Big)\bigg\} \n\\
&& + \epsilon_x \Lambda \bigg\{\!8 \gamma^2\! \cos{(\epsilon)}  \sin{\!\Big(\frac{\epsilon}{2}\Big)}^2\!\! \!+\! \pi \sin{\!\Big(\!\frac{\gamma G \epsilon P_\Lambda}{R}\!\Big)} H_0 \Big(\!\frac{\gamma G \epsilon P_\Lambda}{R}\!\Big) \!\bigg\}\!\bigg].\n
\eea
Here $H_0(x)$ is the Struve function of zeroth order and $\gamma$ is the Immirzi parameter that, for the sake of numerical calculation presented below, we fix it to be approximately 
$0.274$ in consistency with (some) black entropy calculations
in LQG \cite{Agullo:2009eq, Engle:2011vf}. 
The  two quantum parameters in \eqref{heff}, $\epsilon$ and
$\epsilon_x$, are the angular and longitudinal coordinate lengths of the so-called plaquettes, 
cubic cells that are sewn together to make up a graph or a discrete
geometric structure on constant $\tau$ surfaces. 
The graph is such that each 2D hypersurface in a given leaf of our foliation identified by a pair of coordinates is tessellated by a large number of plaquettes, rendering these coordinate lengths very small. More precisely,
due to how the partially gauged fixed LQG kinematical Hilbert space was constructed in \cite{Alesci:2018loi}
%By means of the construction of the partially gauged fixed LQG kinematical Hilbert space in \cite{Alesci:2018loi} 
and the peakedness properties of the coherent states defined on the associated graph structure,
these parameters are related to the semi-classical data as 
%\bea 
% \epsilon& =& \frac{\alpha}{R}\,,\quad \alpha=\sqrt{8 \pi \gamma} \ell_p\sqrt{j_0}\,, \la{epsilon}\\
% \ \epsilon_x & =& \frac{\beta}{\Lambda }\,,\quad \beta=\frac{\sqrt{8 \pi \gamma} \ell_p\, j}{ \sqrt{j_0 }}\,, \la{epsilonx}
%\eea
\be
 \epsilon = \frac{\alpha}{R},\quad \alpha=\sqrt{8 \pi \gamma} \ell_p\sqrt{j_0}; \quad
  \epsilon_x = \frac{\beta}{\Lambda }\,,\quad \beta=\frac{\sqrt{8 \pi \gamma} \ell_p\, j}{ \sqrt{j_0 }}\,, \la{epsilons}
\ee
where $\ell_p = \sqrt{\hbar G/c^3} $ is the Plank length and $j_0$ and $j$ are the  quantum numbers associated respectively with the longitudinal and the angular  links of the coherent states. It is important to emphasize that the above quantum parameters are local quantities that are invariant under a rescaling of global quantities, such as the infrared cut-off $L_0$.
%\footnote{In particular, $\epsilon_x$ can be understood as the ratio between the fiducial length $L_0$ and the number of plaquettes in the $x$ direction. A rescaling of $L_0$ is compensated by a rescaling of the number of plaquettes in the construction of the quantum coherent states used to compute the expectation value of the quantum Hamiltonian constraint, leaving the expression for $\epsilon_x$ independent of this rescaling.}. 
Therefore, the  dynamics derived from the effective Hamiltonian \eqref{heff} with the expressions \eqref{epsilons} for the quantum parameters does not contain any information about $L_0$, rendering our key physical predictions independent of this scale as well.

Let us point out that, 
not surprisingly, in the classical limit where $\hbar \rightarrow 0$, $\epsilon$ and $\epsilon_x$ vanish, the discreteness of quantum geometry  disappears, and the effective Hamiltonian in Eq. \eqref{heff} reduces to its classical value
\bea 
&& \frac{H_{\rm c}}{L_0} = - \frac{G P_R P_\Lambda}{R} + \frac{G \Lambda P_\Lambda ^2}{2 R^2} - \frac{\Lambda}{2 G}.
\eea

The appearance of the Struve function in \eqref{heff} represents the main departure from the minisuperspace quantization models in the previous literature. It is a direct reflection of including ab initio the  fundamental discreteness of quantum geometry on the 2-spheres in the quantum reduced kinematical Hilbert space. The other significant difference, also originating from the $\SU(2)$ holonomies along angular links, is encoded in the term proportional to $\gamma^2$ inside the second curly brackets. This term derives from the Lorentzian part of the LQG Hamiltonian constraint and it contains corrections to all orders in $\epsilon$, while in its minisuperspace counterpart only the leading term is present \footnote{At the same time, this is the only contribution from the Lorentzian term that survives in the homogeneous interior foliation. In all cases considered in this letter, it can be verified that this contribution does not play a dominant role in determining the asymptotic behavior for the effective metrics. Therefore, we expect the ambiguities surrounding the quantization prescription for the Lorentzian piece of the Hamiltonian constraint in the full theory not to have a significant impact on the physical predictions of the effective theory for the interior geometry.}.

In order to distinguish the classical regime, in which general relativity is expected to be a valid description of gravitational field, from the region where quantum effects are expected to become relevant, we rely on 
%a gauge invariant measure of curvature by introducing 
the Kretschmann scalar which for the classical Schwarzschild metric functions given in Eq. \eqref{solc} becomes
% (for the moment we do not commit to a specific choice of lapse)
\be
\mathcal{K}_c:= R_{abcd} R^{abcd}=\frac{3 \ e^{- \frac{3 \tau}{ G m}}}{4 (G m)^4}\,.
\ee  
Quantum gravity heuristics suggest that any effective metric encoding quantum gravity corrections should begin to deviate significantly from the classical one in the regime where curvature becomes (super-) Planckian, namely as $\mathcal{K}_c \gtrsim 1/\ell_p^4$ which happens for times $ \tau \lesssim \tau_\star = (G m/3) \log{[3 \ell_p ^4 /(4 G^4 m^4)]}$. We can thus define the parameter
\be\la{rho}
\rho:={R_c(\tau)}/{R_c(\tau_\star)}
\ee
and expect quantum effects to become dominant at about $\rho\sim1$. 

The quantum parameters $\epsilon$ and $\epsilon_x$ defined in  \eqref{epsilons}  are functions of the effective phase space variables (in the old literature, this corresponds to the so-called $\bar\mu$-scheme \cite{ Bohmer:2007wi}). It follows that the  quantum parameters have non-trivial contributions to the Poisson brackets defining the evolution equations and lead to a very complicated system of ODEs that is difficult to analytically integrate.
We  thus numerically solve the effective Hamilton's evolution equations starting with classical initial data very near the black hole event horizon, namely when $\rho\sim (Gm /\ell_p)^{2/3}$.
We specialize to the following choice of lapse function
 \be \la{N}
N= -\frac{\gamma \epsilon R}{ G m\Big[\sin{\Big(\frac{\gamma G \epsilon P_\Lambda}{R}\Big)} + \frac{\pi}{2} H_0 \Big( \frac{\gamma G \epsilon P_\Lambda}{R}\Big)\Big]}\,,
\ee
as it simplifies the analysis we perform in the asymptotic regime.  This lapse reduces to the classical $N_c$ given above in the limit $\hbar \rightarrow 0$.
In presence of a non-zero $\epsilon$ and for $m$ much larger than the Planck mass $m_p = \sqrt{\hbar c/G}$, the black hole event horizon is still nearly at $\tau = 0$. $\tau$ can be extended all the way to $- \infty$ unless the coordinate system breaks down at a finite value of $\tau$ due to, ${\it e.g}$, reaching a Killing horizon.

The only remaining freedom in specifying the coherent states
is in the choice of spin numbers $j_0$ and $j$. Different choices of $\alpha$ 
 and $\beta$ can appear depending on what one assumes for the quantum numbers $j_0$ and $j$. It turns out  that the qualitative behavior of the solutions to the effective dynamics  depends only the value of the ratio $\eta\equiv \alpha/\beta$ and not on the specific value of the two spin numbers \footnote{Let us remark that the coherent states we used to derive the effective Hamiltonian \eqref{heff} are  peaked already for spin numbers of order 10 (see, e.g., \cite{Livine:2007vk}). In the cases considered here, the bounce happens at an instant of time in vicinity of $\tau_\star$, where the area of the minimal 2-sphere is of order $A\sim (Gm /\ell_p)^{2/3}\ell_p^2$. Therefore, already for black holes of mass $m \sim 10^{12}m_p$, the spin numbers $j\sim j_0\sim 10$ correspond to the minimal 2-sphere being tessallted by $N\sim 10^7$ plaquettes. This is well within our working assumption of a large number of degrees of freedom even in the high curvature regime.
%  and in agreement with the expectations previously supported by other analyses \cite{Han:2016fgh, Gielen:2016uft}. Moreover, from a renormalization flow perspective, a rescaling of the spin in the quantum state can always be understood as a change in the number of placquettes thanks to the peakedness  properties of the Perelemov  SU(2) states, which makes our construction compatible with the description of the non-trivial renormalisation group flow of quantum dynamics provided in \cite{Bodendorfer:2018csn, Bodendorfer:2019wik}.
}. 
We now present results for both when $\eta=1$ and  when $\eta \in (1-\zeta,1+\zeta)$ for $\zeta \ll 1$, as one would expect from the regularity condition on the geometry in the large number of plaquettes limit.
%, as this  seems to require more refined numerical techniques.

\vspace{0.3cm}

{\it The case $\eta=1$}. 
The effective metric functions $R(\tau)$ and $ \Lambda(\tau)$ 
are plotted in FIG. \ref{fig:RL} 
for the choice of $m=10^{12} m_p$ and $j=j_0=10$. The qualitative behavior of the two metric functions remain the same for different values of the mass and spin numbers as long as $m$ is reasonably above the Planck mass (a stellar  mass black hole has $m \approx 10^{38} m_p$).
 \begin{figure}[]
\centering
%\subfloat[A plot of the effective metric function $R$ (blue line) in comparison with the classical metric function $R_c$ (red line). The bounce occurs at $\tau_b= -3.9\times 10^{13}$ (with $\tau_\star=-3.7\times 10^{13}$). The quantities are reported in the unit $m_p = \ell_p = 1$.]
{%
  \includegraphics[width=6.5cm,height=3.5cm]{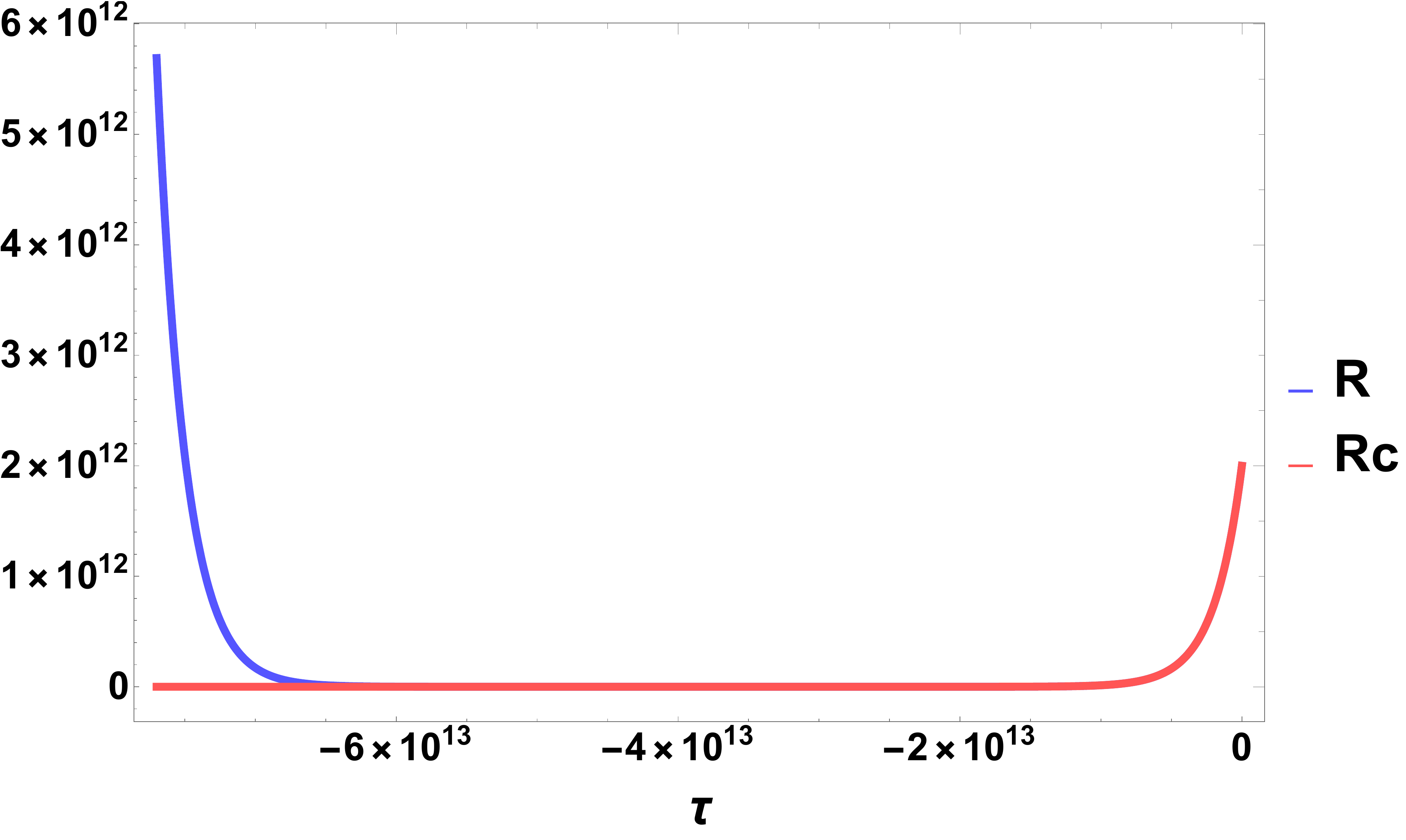}
  }
 
%\subfloat[Plots of the effective metric function $\Lambda$ (blue line) in comparison with the classical metric function $\Lambda_c$ (red line).]
{%
 \includegraphics[width=6.5cm,height=3.5cm]{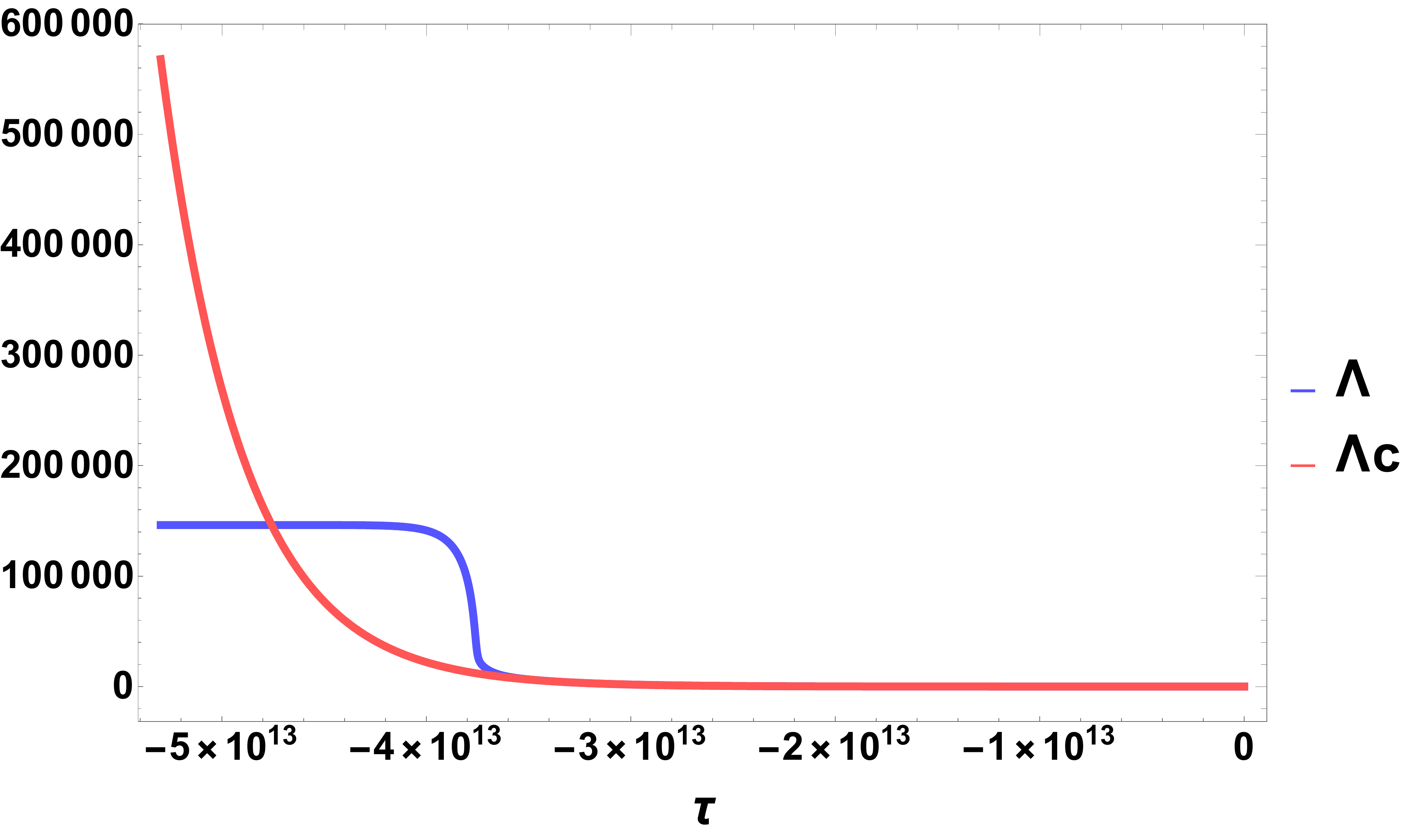}%

}
\caption{Plots of the effective metric functions $R, \Lambda$  (blue lines) in comparison with the classical metric function $R_c, \Lambda_c$ (red lines) for $m=10^{12} m_p$ in the first regime $\alpha=\beta$ for the choice of $j=j_0= 10$. The bounce occurs at $\tau_b= -3.9\times 10^{13}$ (with $\tau_\star=-3.7\times 10^{13}$).
The quantities are reported in the unit $m_p = \ell_p = 1$.}
\la{fig:RL}
\end{figure}
The plots in FIG. \ref{fig:RL} show how the effective metric agrees with the classical one in the low curvature region which spans all the way till $\tau\sim\tau_\star$, where the Planck regime is reached and quantum geometry corrections become dominant, as expected.  In particular, $R$ decreases  with $\tau$ till it reaches a minimum value $R_{\rm min}= R(\tau_b)$. It then bounces and grows  exponentially as $\tau\rightarrow-\infty$. Therefore,  the area of 2-spheres foliating constant $\tau$ surfaces never shrinks to zero, {\it i.e.} the singularity is effectively resolved. The surface $\tau=\tau_b$ marks the transition for the 2-spheres from being trapped in the past ($\tau_b< \tau<0$) to being anti-trapped  in the future ($-\infty<\tau < \tau_b$). In order to determine if a white hole horizon forms in the future of $\tau_b$, we need to analyze the behavior of the second metric function $\Lambda$.

As can be seen in FIG. \ref{fig:RL}, numerical integration suggests that $\Lambda$ approaches a constant non-zero value for
$\tau \ll 0$. An analysis of the Hamilton's dynamical equations confirms this indication. In fact, it can be shown that as $\tau \rightarrow - \infty$ one has $P_R/R\Lambda$ and $2P_\Lambda/R^2$ approaching $-2\pi/\gamma \alpha$, which in turn implies that $\dot R/R \rightarrow -1/2m$ and $\dot \Lambda \rightarrow 0$.
We obtain the following estimates for the metric functions in this limit:
\bea \label{esssol1}
 &&N(\tau)  \rightarrow \text{constant} \sim \frac{ \sqrt{j}  \ell_p}{G m}\,,\n\\  
%\frac{4 \sqrt{2 \pi j} \gamma^{3/2} \ell_p}{\pi H_0 (\pi) G m}, \nonumber\\
 &&\Lambda (\tau)  \rightarrow \text{constant} \sim \left(\!\frac{G m}{\sqrt{j}\ell_p}\!\right)^{\!\!\frac{1}{3}} \,,\n\\
 && R(\tau) \rightarrow \left(\!\frac{j^2 \ell_p ^4}{G m}\!\right)^{\!\!\frac{1}{3}}\!\!\! e^{- \frac{\tau}{ 2 G m}}\,. 
%\sim R_c (\tau_*) e^{- \tau/ 2 G m}\nonumber\\
%&& \hspace{1cm} \sim (G m)^{1/3} \ell_p ^{2/3} e^{- \tau/ 2 G m}.
\eea
%\begin{figure}[H]
%\centerline{ \(
%\begin{array}{c}
%\includegraphics[height=3.5cm]{delta}
%\end{array}\) } 
%\caption{$\delta$}
%\label{fig:delta}
%\end{figure}
Asymptotically, as can be seen from the above equation, the interior geometry becomes 
a product of a 2D pseudo-Euclidean space and a round 2-sphere whose proper area is blowing up exponentially. Note that as $\hbar\rightarrow 0$ the classical singularity reappears. 
It is clear from Eq. \eqref{esssol1} that the interior metric is not asymptotically flat. In fact, the Ricci scalar approaches the non-vanishing asymptotic value $3 / (2 G^2 m^2 N^2) \sim 1/(j_0 \ell_p ^2)$. Some of the other curvature invariants, such as the Ricci squared $R_{ab} R^{ab}$ and the Kretschmann scalar remain non-vanishing but bounded everywhere, while the Weyl squared $C_{abcd} C^{abcd}$ vanishes asymptotically. Interestingly enough, the upper bounds as well as  the asymptotic values of the aforementioned curvature scalars are all independent of the black hole mass $m$ and only carry information about the quantum structure. 
It follows without much difficulty from 
the asymptotic relations \eqref{esssol1} that the interior 
metric is  geodesically complete. The null energy condition is violated for $\tau \lesssim \tau_*$. 

\vspace{0.3cm}

{\it The case $\eta \in (1-\zeta, 1+\zeta)$}. The qualitative behavior of the metric function $R$ in this case is similar to the previous one, with a bounce happening around the critical time $\tau_\star$ followed by a subsequent exponential growth. As for $\Lambda$, it crucially depends on whether $\eta<1$ or $\eta>1$. In fact, performing an asymptotic analysis of the Hamilton's equations indicates that $N$ approaches a constant in the $\tau \rightarrow - \infty$ limit while the other two metric functions assume the following time dependence:
\bea 
&&\Lambda(\tau) \propto e^{\mathcal{F}(\eta) \tau/2 G m}, \ \ \ \  R(\tau)\propto e^{\cos{(\kappa /\eta)} \tau/2 G m}.
\eea
Here $\kappa$ is implicitly defined via 
\be
\eta \sin{(\kappa/\eta)}+ \pi \sin{(\kappa)} H_0(\kappa)/ \big[2 \sin{(\kappa)}+ \pi H_0(\kappa)\big] = 0\,,
\ee
and 
\bea
\mathcal{F}(\eta)  &=& -  \cos{(\kappa/\eta)}\n\\
& +&\frac{ \big[2 \pi H_{-1}(\kappa) \sin^2{(\kappa)} + \pi^2 \cos{(\kappa)} H_0(\kappa)^2\big]}{\big[2 \sin{(\kappa)}+ \pi H_0 (\kappa)\big]^2}\,,
\eea
 with $\kappa = \kappa(\eta)$ and $H_{-1}(x)$  the Struve function of order $-1$. As $\eta$ increases slightly from one, $\kappa$ falls below $-\pi$ and $\mathcal{F}$ becomes negative. On the other hand, as $\eta$ drops slightly below one, $\kappa$ increases from $-\pi$ and $\mathcal{F}$ becomes positive.
%For $1\leq \eta < +\infty$, $-4 < \kappa \leq - \pi$ and $-\infty< \mathcal{F} \leq 0$, 
%while for $0<\eta <1$ we have $ - \pi < \kappa <0$ and $0 < \mathcal{F} <2 $. 
%Additionally,  $\kappa/\eta \rightarrow 0^-$ as $\eta \rightarrow \infty$ whereas $\kappa/\eta \rightarrow -\pi$ as $\eta \rightarrow 0$. Even though we are interested in values of $\eta$ that are in vicinity of one, 
Therefore, the qualitative behavior of $\Lambda$ changes as $\eta$ crosses one due to $\mathcal{F}$ changing sign. Indeed for $\eta = 1 - \zeta$, $\Lambda$ vanishes exponentially while for $\eta = 1+\zeta$ the opposite happens. Nonetheless, the singularity is resolved in both cases. Even in the case when $\Lambda$ vanishes asymptotically, the proper volume of constant $\tau$ surfaces diverges in the same limit since $2\cos{(\kappa/\eta)}+ \mathcal{F}(\eta) < 0$. More precisely, the main curvature invariants, such as the Ricci squared and the Kretschmann scalar remain bounded everywhere. The Ricci scalar becomes $[3+ 2 \mathcal{F}(\eta)^2 + 4 \mathcal{F}(\eta) \cos{(\kappa/\eta)}+3 \cos{(2 \kappa/\eta)}]/[4 G^2 m^2 N^2] \sim 1/(j_0 \ell_p^2)$ asymptotically. As in the special case $\eta=1$, the null energy condition is violated for $\tau \lesssim \tau_*$.
%increases exponentially away from its classical counterpart as the Planckian curvature regime is approached. We numerically find that increasing $\eta$ results in the rates of expansion for $R$ and $\Lambda$ to decrease and increase respectively. $\Lambda$ is plotted in FIG. \ref{fig:L2} for  $\eta = 2$. In this case, too, the Ricci squared and the Kretschmann scalar remain bounded everywhere. They peak around the bouncing time $\tau_b \sim \tau_*$, after which they settle to constant non-zero values.
 
%  \begin{figure}[]
%\subfloat[Regime $\alpha=\beta$, with $j=j_0=10$.]
%{  \includegraphics[clip,width=0.75\columnwidth,center]{L2}%
%}
%
%\subfloat[Regime $\alpha>\beta$, with $j=10, j_0=20$.]{%
%  \includegraphics[clip,width=0.85\columnwidth]{delta2-both}%
%}
%\caption{Metric function $ \Lambda$ for $m=10^{12} m_p$ in the second regime $\alpha>\beta$ for the choice of $j= 10, j_0=20$.}
%\la{fig:L2}\vspace{-0.3cm}
%\end{figure}

\vspace{0.3cm}

{\it Large quantum effects.} In the minisuperspace quantization literature,
the  $\bar\mu$-scheme was first applied to the Schwarzschild interior in \cite{ Bohmer:2007wi}. 
There it was noted that large quantum effects near the classical 
event horizon appear, contrary to the  expectation of such corrections to be suppressed in the low curvature regime. However, this undesirable feature is absent in our effective description. 
Let us elaborate further on this important point.
%There the effective metric was shown to be
%%found also through a combination of numerical and analytical techniques showed an 
%a Nariai type Universe in the asymptotic limit, with the effective $R$ undergoing damped oscillations until a fixed finite value was reached as $\tau\rightarrow-\infty$, while $\Lambda$ blew up exponentially in 
%%this asymptotic regime
%the same limit. 
%%In addition to the clearly opposite behavior of our
%Aside from obtaining distinct qualitative behavior for 
%the metric functions in the asymptotic limit,
% asymptotic effective solutions \eqref{esssol1} obtained with the first prescription \eqref{epsilons} for the quantum parameters, 
%our effective description circumvents the physically undesirable large quantum effects near the classical 
%event horizon that appeared in the $\bar\mu$-scheme of \cite{ Bohmer:2007wi}, and more generally in the polymer quantization approach. Let us elaborate further on this important point.
If we quantify  quantum geometry corrections by the ratio
$
\delta:=\left(\mathcal{K}/\mathcal{K}_c\right)^{1/4}
$
of the effective over the classical  Kretschmann scalar, large quantum effects 
necessitate significant deviations from unity for this parameter. 
%It can be verified numerically that, for all the cases  $\eta=1, \eta>1, \eta<1$, the quantity $\delta$ deviates from 1 only as we enter the deep Planckian regime $\rho\sim 1$, indicating that there are no large quantum effects.
In FIG. \ref{fig:rho} we plot this ratio  as a function of the parameter $\rho$ defined previously in \eqref{rho} for the case $\eta=1$, but a similar behavior is found also for the other two cases   $ \eta>1, \eta<1$. As shown, there are no large quantum effects, as the effective geometry deviates significantly from the classical one only in the regime where the curvature becomes Planckian, namely as $\rho\sim 1$.
 \begin{figure}[h]
 \vspace{-0.35cm}
%\subfloat[Regime $\alpha=\beta$, with $j=j_0=10$.]
{  \includegraphics[width=.4\textwidth,center]{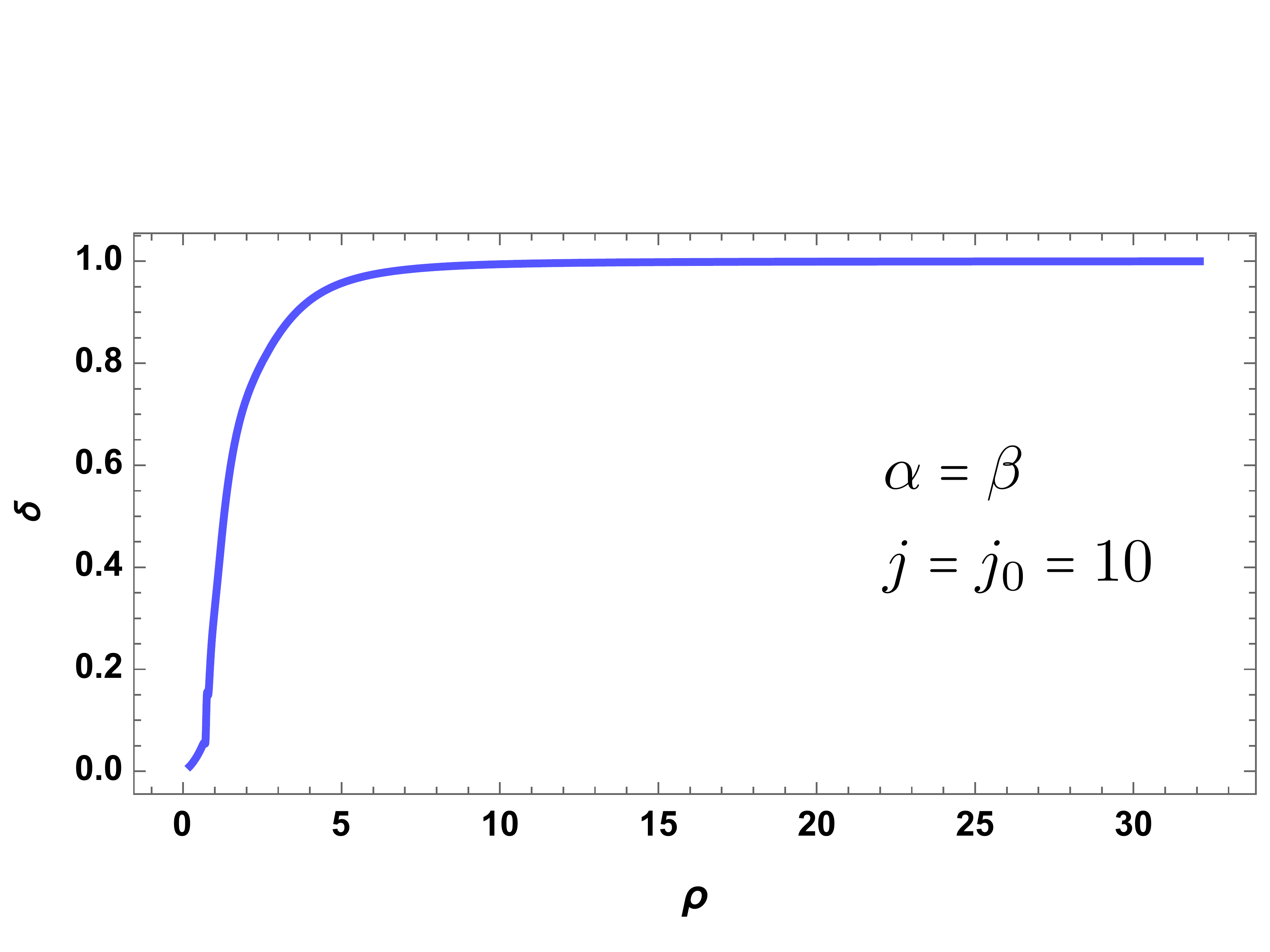}%
} \vspace{-0.5cm}
%
%\subfloat[Regime $\alpha>\beta$, with $j=10, j_0=20$.]{%
%  \includegraphics[clip,width=0.85\columnwidth]{delta2-both}%
%}
\caption{The plot shows how, in the case $\eta=1$, the ratio between the effective and the classical  Kretschmann scalar deviates from 1 only as we enter the deep Planckian regime $\rho\sim 1$. Here we set  $m=10^{12} m_p$, but the same conclusion holds for much larger values of $m$, as well as in the other two cases $\eta>1, \eta<1$.}
\la{fig:rho}\vspace{-0.3cm}
\end{figure}

% \begin{figure}[H]
%\centering
%\subfloat[Regime $\alpha=\beta$, with $j=j_0=10$.]{%
%  \includegraphics[clip,width=0.85\columnwidth]{delta1-both}%
%}
%
%\subfloat[Regime $\alpha>\beta$, with $j=10, j_0=20$.]  {%
% \includegraphics[clip,width=0.85\columnwidth]{delta2-both}%
%}
%\caption{The plots show how, in the two regimes analyzed, the ratio between the effective and the classical  Kretschmann scalar deviates from 1 only as we enter the deep Planckian regime $\rho\sim 1$. In both plots we set  $m=10^{12} m_p$, but the same conclusion holds for much larger values of $m$.}
%\la{fig:rho}
%\end{figure}

\vspace{0.3cm}
{\it Discussion of results.} In this letter we  investigated the predictions of an effective quantum theory applied to the Schwarzschild black hole interior that was derived for the first time  from the full LQG framework. While ambiguities related to the choice of quantization prescription for the full Hamiltonian constraint still remain, 
%this does not allow us  to make  general statements about the full theory yet, as we haven't analyzed in detail how all possible corrections reflecting different  quantization prescriptions of the full Hamiltonian constraint affect the effective solutions, 
our analysis represents an advancement in the field and an important step in the right direction. Despite confirming some key features of the minisuperspace quantization models, our predictions show a number of significant differences as well.  

%, providing an extension of the classical Penrose diagram for the interior geometry where a null conformal boundary is reached. 
More precisely, for all coherent states parameters considered here, the effective metric function $R$ has the same qualitative behavior; it follows the classical trajectory until the Planckian curvature regime is reached, where quantum geometry effects generate a bounce after which an exponential expansion follows. All  curvature scalars are finite at any time  with mass-independent upper bounds, indicating that the classical singularity is resolved.
%Therefore, our analysis provides the first robust evidence that LQG resolves the black hole singularity.
% From Sina: I think this sentence is repetative.

However, the three different possibilities for the ratio between the two quantum spin numbers result in qualitatively different post-bounce effective geometries, as described in the main body of the letter. These differences are due to the behavior of the effective metric function $\Lambda$ after the bounce.
% and implied also by the different asymptotic behavior of the second effective metric function $\Lambda$.
In the first case where $\eta=1$, $\Lambda$ ceases to change appreciably around the bouncing time and reaches a constant asymptotic value.  In the second case where  $\eta>1$, $\Lambda$ grows exponentially around the bouncing time at a rate that increases for larger values of $\eta$.  Finally for  $\eta<1$, $\Lambda$ decreases exponentially after the bounce.
However, in all these cases $\Lambda$ remains non-zero %again (recall that $\tau=0$ corresponds to the black hole horizon where $\Lambda=0$ initially) 
at any finite time $\tau <0$, implying that {\it no white hole horizon forms after the bounce}.

Let us conclude this discussion with some remarks on the shortcomings of our analysis. 
Firstly, the issue of off-shell anomaly freedom of the effective dynamics is left open in our analysis, as in most minisuperspace quantization approaches in the previous literature (see, however, \cite{Bojowald:2018xxu, BenAchour:2018khr} for an alternative approach). This is an important consistency check, as the physical relevance of the effective line element crucially depends on the closure of  the hypersurface-deformation algebra. A possible, if not likely, outcome is that the radial diffeomorphism constraint also needs to be modified. While this may not have relevant implications for the interior homogeneous foliation considered here since the shift vector can always be eliminated by changing the $x$ coordinate, a consistent derivation of effective diffeomorphism constraint becomes crucial when concurrently solving for both the interior and exterior regions. In this case, the inhomogeneity of  spacetime prevents one from setting the shift vector to zero. This is an ongoing area of research.

Secondly, at this stage, our derivation of the effective dynamics is based on a fixed graph structure. It would be very important to relax this restriction by performing a sum over plaquettes on all three orthogonal planes of our cuboidal decomposition. A crucial ingredient for this procedure is the choice of weight for each graph.  In the simpler cosmological setting \cite{Alesci:2017kzc}, a somehow ad hoc combinatorial factor was chosen motivated by the statistical counting of microstates compatible with macroscopic configurations. In the black hole case, previous experience with entropy counting may lend important guidance for a physically better motivated choice. Whether the effective solutions obtained here are robust after inclusion of new quantum corrections encoding  fluctuations in the number pf plaquettes is an important question we are working on to address in the near future.

In this regard, let us also point out how an alternative approach to the construction of a quantum black hole  geometry within the full theory, preserving the combinatorial information of the graph structure, has been developed in the past few years using the operatorial formulation of group field theory \cite{Oriti:2015qva, Oriti:2015rwa, Oriti:2018qty}. This framework provides a description of homogeneous continuum geometries by relying on the use of condensate states whose wave function is peaked on a few  global observables, hence differing from the coherent states used in \cite{Alesci:2018loi} that instead peak on the classical values of local geometrical quantities  for individual links of the underlying graphs. While this feature of group field theory condensates  allows to retain better control on the sum over (a given set of) graphs and the semiclassicality nature of the states by guaranteeing small fluctuations for some global geometrical operators, it also has its shortcomings. In particular, the difficulty in defining local operators introduces a higher level of ambiguity and technical difficulty in defining the dynamics of the theory. Our hope is that the two approaches can lead to a fruitful cross-fertilization of ideas and techniques, that could eventually elucidate more in detail the interpretation of dynamics implementation as a refinement process.

\vspace{0.3cm}
{\it  Comparison with existing literature.}
Recent calculations in the  LQC minisuperspace quantization  literature  have indicated that the post-bounce anti-trapped region admits 
%a boundary that can be interpreted as 
a white hole horizon \cite{Ashtekar:2018lag, Ashtekar:2018cay}.
 A comparison between these observations and our results
is merited in order to highlight the key role played by the quantum geometry corrections that are accounted for in our effective Hamiltonian \eqref{heff}.
%The regularization prescription adopted in  \cite{Ashtekar:2018lag, Ashtekar:2018cay} is a hybrid between the $\bar \mu$-scheme  and the $\mu_0$-scheme, where the quantum parameters are fixed to be  constants along the effective dynamical trajectories. 
In the regularization prescription adopted in  \cite{Ashtekar:2018lag, Ashtekar:2018cay} the quantum parameters are fixed to be  constants along the effective dynamical trajectories. 
The accuracy of the evolution equations derived in \cite{Ashtekar:2018cay} through this scheme has been questioned in
\cite{Bodendorfer:2019xbp} (see also \cite{Bouhmadi-Lopez:2019hpp, Bojowald:2019dry} for issues with asymptotic flatness and  general covariance affecting the model constructed in  \cite{Ashtekar:2018lag, Ashtekar:2018cay}). However, since our goal here is to investigate whether the quantum corrections coming from a full theory Hamiltonian constraint reproduce the qualitative behavior of the effective interior metric derived in this hybrid scheme through point holonomies techniques, we analyze the case where the quantum parameters do not contribute to the Poisson brackets determining the evolution  equations. To this end, 
instead of relying on the questioned approach based on the Dirac observables, 
we simply start by considering constant quantum parameters. 
%  then relate these parameters to the classical black hole mass $m$ by imposing a physical condition on the minimal 2-sphere area.
We then relate these parameters to the classical black hole mass by imposing a physical condition on the minimal 2-sphere area in order to arrive at a final expression that closely resembles the one used in \cite{Ashtekar:2018cay}.

%%%%%%%%%%%%%%%%%%%%%%%%%%%%%%

%%%%%%%%%%%%%%%%%%%%%%%%%%%%%%

%Before making a specific choice, let us mention a few  results that follow in this analysis.
%Firstly, it can be verified that
%\be \label{do}
%\mathcal{M} \equiv  \frac{R^2 \sin{\Big(\frac{\gamma G \epsilon_x }{R^2} [P_R R - P_\Lambda \Lambda]\Big)}}{\gamma G \epsilon_x}
%\ee
%is a Dirac observable, namely $\{\mathcal{M}, H_{\rm eff} \}\approx 0$ (where $\approx$ denotes evaluation on the constraint surface), that reduces to the
%classical gravitational mass $m$ in the limit $\hbar \rightarrow 0$.

In the case of constant $\epsilon$'s, the convenient choice \eqref{N} 
for the lapse
function 
%(where we can now replace $m$ with $\mathcal{M}$)
allows for the explicit solution 
%\footnote{Here $m$ can also be understood as a Dirac observable of the effective dynamics generated by constant $\epsilon$'s  that reduces to the
%classical  gravitational mass in the limit $\hbar \rightarrow 0$. In fact, the two are basically the same as long as $m \gg m_p$.}
%In fact, $m$ would be almost the same as the classical Schwarzschild mass  so long as the spacetime curvature is sub-Planckian near the event horizon, or equivalently as long as $m \gg m_p$. 
%As the system of ODEs in our effective dynamics is not integrable, we think it is unlikely that there be a second Dirac observable, as one might have suspected from the discussion in \cite{Bodendorfer:2019cyv}.}
\be \la{R}
R(\tau) = 2 Gm e^{\frac{\tau}{2 Gm}} \sqrt{1 + \frac{\gamma^2 \epsilon_x ^2 }{64 G^2m^2} e^{-\frac{2\tau}{ Gm}}}\,.
\ee
%\be \la{R}
%R(\tau) = 2 G \mathcal{M} e^{\frac{\tau}{2 G \mathcal{M}}} \sqrt{1 + \frac{\gamma^2 \epsilon_x ^2 }{64 G^2 \mathcal{M}^2} e^{-\frac{2\tau}{ G \mathcal{M}}}}\,.
%\ee
Here $m$ can also be understood as a Dirac observable of the effective dynamics generated by constant $\epsilon$'s  that reduces to the
classical  gravitational mass in the limit $\hbar \rightarrow 0$. In fact, the two are basically the same as long as $m \gg m_p$.
 In the limit $\hbar \rightarrow 0$, we recover 
 the classical metric function given in Eq. \eqref{solc}.
It is clear from Eq. \eqref{R} that the aerial coordinate
reaches a minimum value 
$
R_{\rm min} (\tau_b) = \sqrt{\gamma \epsilon_x Gm} 
$
%\be 
%R_{\rm min} (\tau_b) = \sqrt{\gamma \epsilon_x G \mathcal{M}} 
%\ee
at the moment $\tau_b = Gm \log{\big(\gamma \epsilon_x  / 8 Gm\big)}$.
%$\tau_b = G \mathcal{M} \log{\big(\gamma \epsilon_x  / 8 G \mathcal{M}\big)}$. 
This is to say that the area of
the concentric 2-spheres that foliate 
constant $\tau$ surfaces never shrinks to zero, evidencing how the quantum corrections eliminate the classical singularity  and replace it with a cosmological bounce
at $\tau = \tau_b$. 

Let us now provide a physical argument to relate the quantum parameters to the black hole mass $m$.
% to be constants along the dynamical trajectories. 
 A natural choice would be to equate $R_{\rm min}(\tau_b)$  with the classical aerial radius when the curvature is Planckian, namely 
$
R_{\rm min} (\tau_b)= R_c (\tau_\star)\,.
$
This immediately yields %$\epsilon_x=3^{1/3}2^{4/3} \ell_p ^{4/3}/(\gamma (Gm)^{1/3})$
\be\la{epsilonx-merda}
\epsilon_x=\frac{3^{1/3}2^{4/3} \ell_p ^{4/3}}{\gamma (Gm)^{1/3}}\,.
\ee
%\be\la{epsilonx}
%\epsilon_x=\frac{3^{1/3}2^{4/3} \ell_p ^{4/3}}{\gamma (G \mathcal{M})^{1/3}}\,,
%\ee
% where we have implicitly assumed that the effective geometry
%approaches the classical one in proximity of the black hole event horizon, ${\it i.e.}$ $\mathcal{M}$ is the Schwarzschild gravitational mass. This is correct so long as
%the spacetime curvature is sub-Planckian near the event horizon, or equivalently as long as $\mathcal{M} \gg m_p$.
We can then fix $\epsilon$ by demanding that at the bouncing time $j_0=1/2$ (for different values of $j_0$ the qualitative results remain unchanged), namely 
%$\epsilon =\sqrt{ {4 \pi \gamma \ell_p ^2}/{(\gamma \epsilon_x Gm)}}={2^{1/3}\sqrt{\pi \gamma} \ell_p ^{1/3}}/ { (3^{1/6}( Gm)^{1/3})}$,
\bea\la{epsilon-merda}
\epsilon =\sqrt{ \frac{4 \pi \gamma \ell_p ^2}{\gamma \epsilon_x Gm}}=\frac{2^{1/3}\sqrt{\pi \gamma} \ell_p ^{1/3}} { 3^{1/6}( Gm)^{1/3}}\,,
\eea
%\be\la{epsilon}
%\epsilon =\sqrt{ \frac{4 \pi \gamma \ell_p ^2}{\gamma \epsilon_x G \mathcal{M}}}=\frac{2^{1/3}\sqrt{\pi \gamma} \ell_p ^{1/3}} { 3^{1/6}( G \mathcal{M})^{1/3}}\,,
%\ee
where we used Eq. \eqref{epsilons}. 
Therefore, as in the regularization scheme of \cite{Ashtekar:2018cay}, the two quantum parameters acquire a dependence on  the gravitational mass as a power of $-1/3$.
Numerical investigations of Eq. \eqref{peq} show that all commonly used spacetime curvature invariants, such as the Kretschmann scalar, remain bounded for all times $- \infty< \tau< 0$, solidifying the resolution of black hole singularity in this model. Moreover, we numerically confirm a feature found in  \cite{Ashtekar:2018lag} that suggests that the upper bounds for these curvature invariants are mass independent as long as $m \gg m_p$.  

However, the similarities end here.
 In particular, by omitting the  corrections coming from the 2-sphere quantum geometry due to using point holonomies for the angular directions, the effective dynamics of \cite{Ashtekar:2018lag} predicts the appearance of a white hole horizon at a finite instant of time $\tau$. As emphasized in the beginning of this letter, the effective dynamics derived in  \cite{Alesci:2018loi}  
accounts for precisely these ignored degrees of freedom that are present in the full theory. 
%In the expression of the effective Hamiltonian \eqref{heff}, the quantum effects associated with the discreteness of quantum geometry on the 2-spheres are encoded in the appearance of the Struve function of zeroth order, instead of a Sine function. 
This modification has drastic implications for predictions of the theory. In fact, our quantum-corrected interior geometry prevents the formation of a white hole horizon at any time.
More precisely, 
%while in this case too it turns out to be difficult to find analytical expressions for all of the metric functions,
 the following asymptotic
estimates that are valid in the limit $\tau \rightarrow - \infty$ can be obtained:
\bea \la{essol}
&& N(\tau) 
 \sim  \frac{\epsilon  R}{Gm}
\!  \sim \! \frac{\epsilon \  \epsilon_x}{Gm} e^{-\frac{\tau }{2 Gm}} \,,\n\\
&&\Lambda(\tau) \!\sim\!  \frac{\epsilon Gm}{\epsilon_x} e^{-\frac{\tau \mathcal{L}(\epsilon)}{2 Gm}}\,,\n\\
&& R(\tau)\! \sim  \!\epsilon_x e^{-\frac{\tau }{2 Gm}}.
\eea
%\bea \la{essol}
%&& N(\tau) \sim  \frac{\epsilon  R}{G \mathcal{M}} \sim  \frac{\epsilon \  \epsilon_x}{G \mathcal{M}} e^{-\tau /2 G \mathcal{M}} , \nonumber\\
%&& \Lambda(\tau) \sim \epsilon \frac{G \mathcal{M}}{\epsilon_x} e^{-\tau \mathcal{L}(\epsilon)/2 G \mathcal{M}}, \nonumber\\
%&& R(\tau) \sim  \epsilon_x e^{- \tau/ 2G \mathcal{M}}.
%\eea
Here 
$
\mathcal{L}(\epsilon) = {2 \gamma^2 \epsilon^2 \big[2+ \pi H_{-1} (\pi)\big]}/{ \pi^2 H_0 ^2 (\pi)} + \mathcal{O}(\epsilon^3)\,,
$
which is numerically found to be greater than zero for $\epsilon \ll 1$, but significantly suppressed compared to unity. 
This asymptotic behavior of $\Lambda$ is confirmed by the numerical solution, that shows how
 %as shown in FIG. \ref{fig:Lmerda}. As can be seen from the plot, 
 $\Lambda$ reaches a maximum value around the bouncing time, then decreases until a minimum value greater than zero is reached, and  then  grows very slowly as $\tau \rightarrow - \infty$. Therefore, $\Lambda$ never goes to zero after the initial instant of time corresponding to the black hole horizon.
%\begin{figure}[h]
%\centering
%  \includegraphics[width=.35\textwidth,center]{Lmer}
%  \caption{Plot of the effective metric functions $\Lambda$ (blue line) in comparison with the classical metric function $\Lambda_c$ (red line)  for $m=10^{12} m_p$ in the analog of the $\mu_0$-scheme; the bounce occurred at $\tau_b\sim -3.763\times 10^{13}$.}
%%\caption{Plot of the effective metric functions $\Lambda$ (blue line) in comparison with the classical metric function $\Lambda_c$ (red line)  for $m=10^{12} m_p$ in the analog of the $\mu_0$-scheme.  The zoom shows the behavior of $\Lambda$ after the bounce occurred at $\tau_b\sim -3.763\times 10^{13}$.}
%\la{fig:Lmerda}\vspace{-0.3cm}
%\end{figure}

%Since all metric functions diverge at most as fast as $R$ in this limit, we can conformally complete the interior metric by 
%the conformal factor $\Omega = G \mathcal{M}/R$, from which it follows that in this case too
%the interior spacetime is endowed with a null conformal boundary. 
Let us conclude by remarking that,
as in the previous case, the interior metric is geodesically complete but not asymptotically flat. However, now the components of the Riemann tensor computed in an orthonormal frame vanish asymptotically. 
%Similarly, all curvature invariants, such as the Ricci scalar and the Kretschmann scalar, vanish asymptotically as well. 
%The Penrose diagram for this interior region is identical to the one provided in FIG. \ref{penrose1}. 
Moreover, the weak, strong and dominant energy conditions are violated only around the bounce $\tau \sim \tau_b \sim \tau_*$, while they are satisfied as $\tau$ gets more negative.

\vspace{0.3cm}
{\it Acknowledgments.}
We are grateful to A. Ashtekar and J. Olmedo for helpful discussions. 
E. Alesci and S. Bahrami are supported in part by the NSF grant PHY-1505411, the Eberly research funds of Penn State, and the Urania Stott fund of Pittsburgh Foundation.
Research at the Perimeter Institute  is supported in part by the Government of Canada through NSERC and by the Province of Ontario through MRI.

%\bibliography{Reference}% Produces the bibliography via BibTeX.

\end{document}